%% file: ew_fit_gfitter.tex
\documentclass[a4paper,11pt]{article}
\usepackage{pos}
\usepackage{amsmath} 
\usepackage{xspace}  
\usepackage{wrapfig} 

\input definitions

\title{Status of the global electroweak fit with Gfitter in the light of new precision measurements}
\ShortTitle{The global electroweak fit with Gfitter}

\author[a]{Johannes Haller}
\author[b]{Andreas Hoecker}
\author*[c]{Roman Kogler}
\author[c]{Klaus M\"onig}
\author[b]{J\"org Stelzer}

\affiliation[a]{Institut f\"ur Experimentalphysik, Universit\"at Hamburg, Hamburg, Germany}
\affiliation[b]{CERN, European Organization for Nuclear Research, Geneva, Switzerland}
\affiliation[c]{Deutsches Elektronen-Synchrotron DESY, Hamburg and Zeuthen, Germany}

\emailAdd{roman.kogler@desy.de}

\abstract{
  We present results from the global electroweak fit to precision measurements in the Standard Model (SM). The fit uses the latest theoretical calculations for observables on the $Z$ pole and the $W$ boson mass, yielding precise SM predictions for the effective weak mixing angle and the masses of the $W$ and Higgs bosons, as well as the top quark. We study the impact of the latest measurements on the fit and provide comparisons of the resulting predictions for individual observables with recent measurements.
}

\FullConference{%
  41st International Conference on High Energy physics - ICHEP2022\\
  6-13 July, 2022\\
  Bologna, Italy
}

\begin{document}
\maketitle

\section{Introduction}

Electroweak precision data from LEP and SLD~\cite{ALEPH:2005ema, Schael:2013ita}, 
together with precise measurements at low energy~\cite{Davier:2017zfy},  
and direct measurements of the parameters of the Standard Model (SM) 
by the Tevatron and LHC experiments, 
can be used to test the internal consistency of the theory. 
In the past, the electroweak (EW) fit was able to predict the masses of the top quark \mt 
and the Higgs boson \MH. With their discoveries, the EW sector of the 
SM is complete and the fit is overconstrained. This allows for consistency tests 
with unprecedented precision. 
An integral part of these tests are calculations which match or exceed the experimental 
measurements in precision. In the last years, the full two-loop EW 
contributions for \PZ boson production and decay have been calculated~\cite{Dubovyk:2016aqv, Dubovyk:2018rlg}.  
These calculations conclude a long effort in theoretical physics, which 
have resulted in theoretical uncertainties smaller than the experimental ones 
in all observables entering the EW fit. 
Comparisons between measurements and predictions of key observables like the mass of the \PW boson 
\MW and the fermionic effective EW mixing angle \sinfeff allow for concise statements about the 
validity of the SM and can show shortcomings of the theory. 

In this contribution, we update our previous results~\cite{Baak:2014ora, Haller:2018nnx} 
with new measurements of \MW, \mt, and \sinleff. In particular, we investigate the 
effect of the most recent \MW measurement by the CDF Collaboration~\cite{CDF:2022mw}. 

\section{Measurements and calculations}

\begin{figure}
\centering
\includegraphics[width=0.48\textwidth]{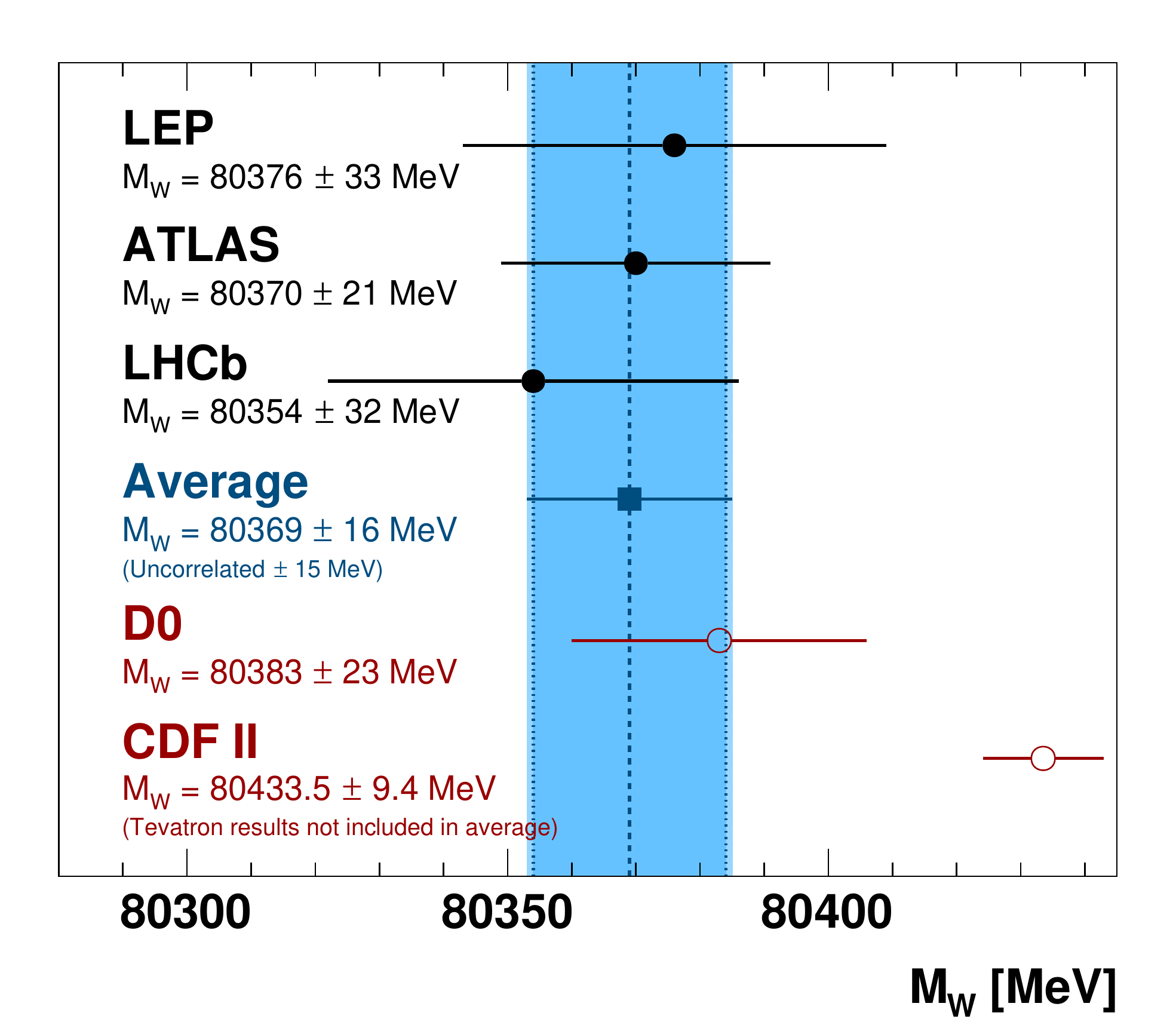}\hspace{0.1cm}
\includegraphics[width=0.48\textwidth]{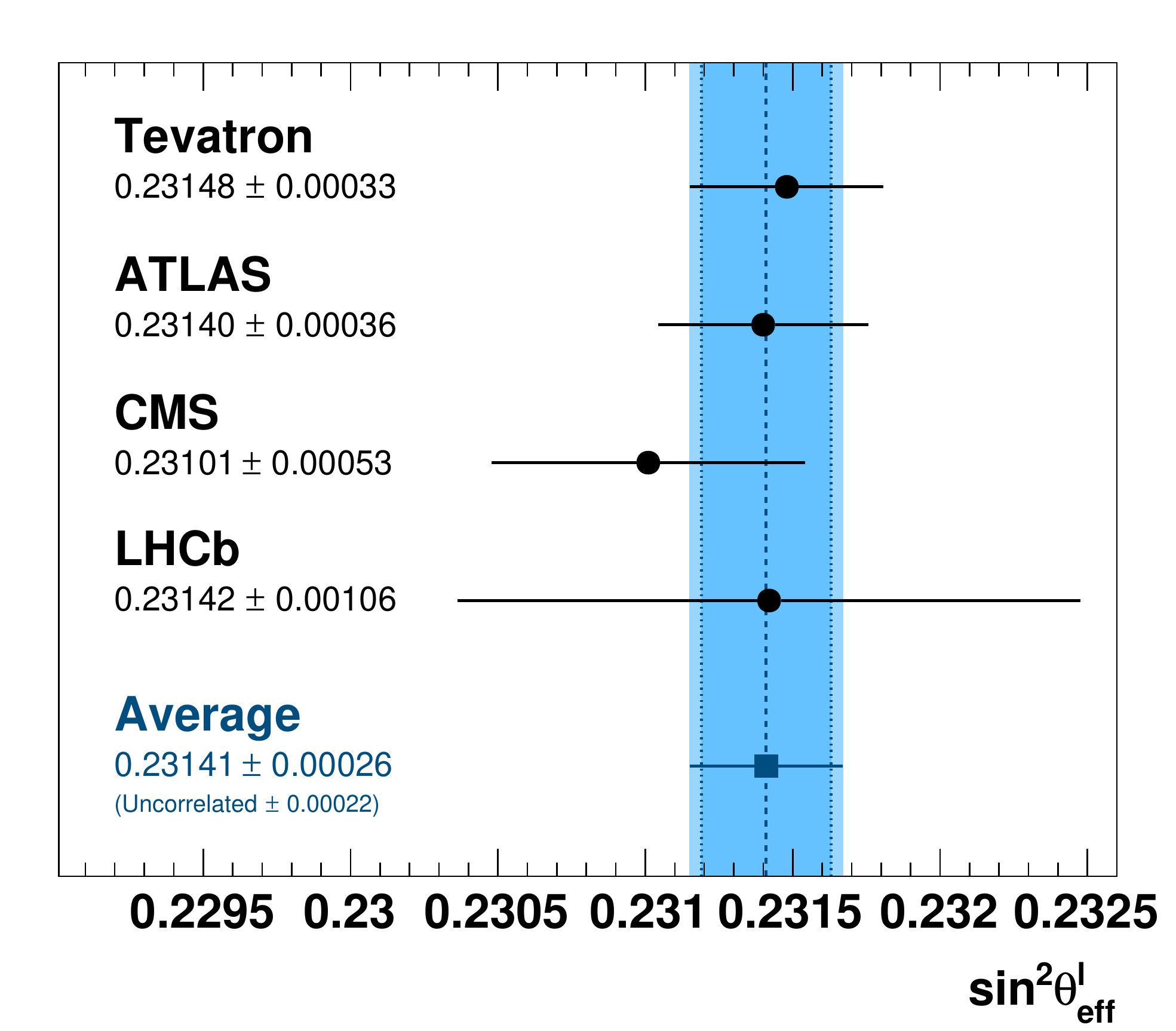}
\caption[]{Individual measurements and combination of the $W$ boson mass (left) 
and $\sinleff$ (right). The red measurements by the D0 and CDF Collaborations 
are not included in the combination of \MW. }
\label{fig:combinations}
\end{figure}

Since our previous results~\cite{Haller:2018nnx}, new measurements of \MW have been 
performed by the LHCb~\cite{LHCb:2021bjt} and CDF~\cite{CDF:2022mw} 
Collaborations. LHCb reports a value of $80354 \pm 32 \mev$, which is 
in agreement with the LEP average~\cite{Schael:2013ita} of $80376 \pm 33\mev$ and has a comparable uncertainty. 
The value obtained by CDF using the full Tevatron Run II dataset is $80433.5 \pm 9.4\mev$, which is 
the most precise single measurement to date. However, it disagrees by about 2.7 standard deviations ($\sigma$) 
with the second most precise measurement by ATLAS, $\MW = 80370 \pm 21\mev$~\cite{Aaboud:2017svj}. 
The exact size of the disagreement depends on the assumed correlations between these measurements, 
where $2.7\sigma$ is the minimum deviation obtained for no correlations.
Since the LEP, LHCb and ATLAS measurements agree within $1\sigma$ of their uncertainties, 
we build an \MW average from these three. We do not consider an older measurement by the D0 Collaboration, 
because effects from the signal modelling and parton distribution functions (PDFs) 
studied by the LHC EW Working Group~\cite{CERN-LPCC-2022-06} 
in the context of the CDF measurement could affect this measurement as well. 
The resulting average is $\MW = 80369 \pm 16\mev$ with a $\chi^2$ of 0.28 for 2 degrees of freedom (dof). 
This average is compared to the individual measurements in Figure~\ref{fig:combinations} (left). 
The result is robust against changes in the correlations of the modelling uncertainties of the 
ATLAS and LHCb measurements. It is smaller by 10\mev than the value of \MW used in our previous fit. 
Our average disagrees with the CDF Run II value by 3.5 to $4\sigma$, which corresponds to $p$ values 
between $5\cdot10^{-4}$ and $5\cdot10^{-5}$, where the range reflects different 
choices of correlations.

For \mt, we use the same combination of ATLAS and CMS measurements 
as in our previous fit~\cite{Haller:2018nnx}, which results in $\mt = 172.47 \pm 0.46\gev$. 
We add an uncertainty of 0.5\gev in \mt to account for 
a potential ambiguity when translating the measured value into the 
top quark pole mass. 

For \sinleff, we include the direct measurement by the LEP Collaborations~\cite{ALEPH:2005ema} and 
a combination from measurements at hadron colliders. 
The combination is made from the measurements by ATLAS~\cite{ATLAS-CONF-2018-037}, 
CMS~\cite{CMS:2018ktx}, LHCb~\cite{LHCb:2015jyu}, and the combined value from CDF and D0~\cite{CDF:2018cnj}. 
We correlate the PDF uncertainties fully between the ATLAS and CMS measurements, and 
we assume a correlation of 50\% between the Tevatron and ATLAS/CMS PDF uncertainties, 
as well as between LHCb and ATLAS/CMS. In addition, we use a correlation of 30\% between the 
Tevatron and LHCb PDF uncertainties. The resulting value is $\sinleff = 0.23141 \pm 0.00026$, 
shown in Figure~\ref{fig:combinations} (right). The combination shows good 
compatibility of the individual measurements with  
$\chi^2/\mathrm{dof} = 0.74/3$, resulting in a $p$ value of 0.86. 

Besides electroweak precision data from LEP and SLD~\cite{ALEPH:2005ema}, 
other experimental inputs to the fit are the hadronic contribution to the electromagnetic 
coupling strength~\cite{Davier:2017zfy}, \MH~\cite{Aad:2015zhl} and the masses
of the $c$ and $b$ quarks~\cite{Workman:2022ynf}.

An integral part of the EW fit are precise calculations. 
The theoretical higher-order calculations 
used here are the same as in our previous publication~\cite{Haller:2018nnx}. Most notably, 
we use the two-loop EW contributions for \PZ boson production and 
decay~\cite{Dubovyk:2016aqv}, where we leave the inclusion of the latest calculations 
of bosonic contributions~\cite{Dubovyk:2018rlg} to future work. 
For \sinfeff, we use the parametrisation of Ref.~\cite{Dubovyk:2016aqv}, 
and the prediction of \MW is obtained from Ref.~\cite{Awramik:2003rn}. 
The width of the $W$ boson is known up to one-loop order in the EW interaction, 
where we use the parametrisation given in Ref.~\cite{Cho:2011rk}.
Theoretical uncertainties reflect the size of unknown higher order contributions. 
Since these are difficult to estimate, a reliable consistency test of the SM is 
only obtained if the theoretical uncertainties are small compared to the experimental uncertainties. 
We introduce a free parameter for each theoretical uncertainty in the EW fit, 
and find that the impact of these parameters is small.

\section{Results of the electroweak fit}

The fit converges on a minimum $\chi^2$ value of 16.62 for 15 dof, 
corresponding to a $p$ value of 0.34. 
The individual deviations of the input values from the predictions at the best-fit point 
given in units of 
\begin{wrapfigure}{r}{0.5\textwidth}
    \centering
    \includegraphics[width=0.48\textwidth]{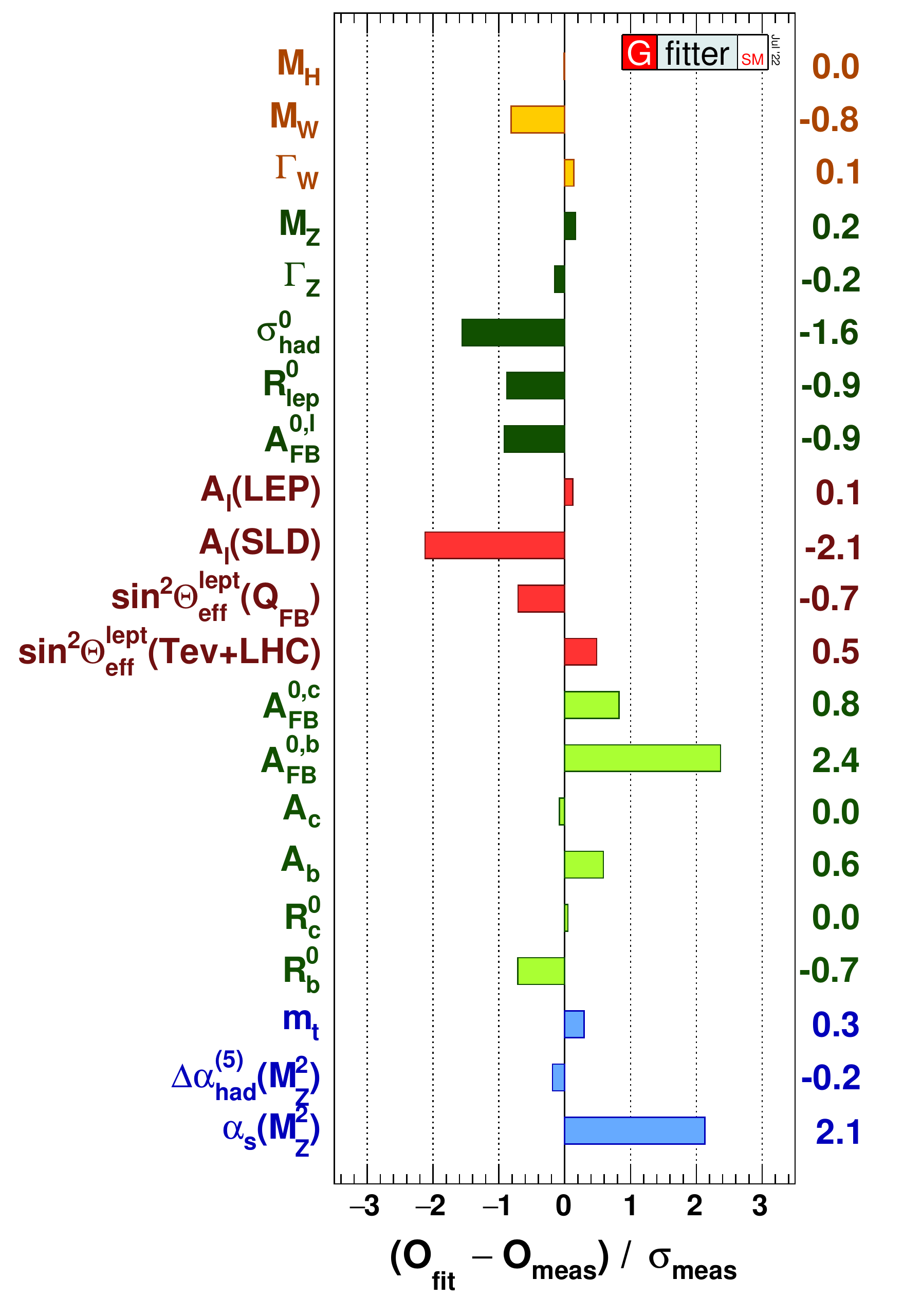}
  \caption{Pull values of the fit, defined as deviations between measurements 
  and predictions evaluated at the best-fit point, divided by the experimental uncertainties. \label{fig:pulls}}
\vspace{-0.3cm}
\end{wrapfigure}
the measurement uncertainty (pull values), are shown in Figure~\ref{fig:pulls}. The largest contribution to the 
$\chi^2$ originates from the forward-backward 
asymmetry from $b$ quarks, $\Afb{b}$, which 
shows a deviation of $2.4\sigma$ from the SM prediction. 
The leptonic left-right asymmetry $A_\ell$ from SLD has a deviation 
of $-2.1\sigma$. These two effects are unchanged with 
respect to our previous results. 
The best-fit value of the strong coupling strength at the mass of the \PZ boson, 
$\asZ$, is $0.1198\pm 0.0029$. In Figure~\ref{fig:pulls}, this is compared to the PDG value 
of $0.1179\pm0.0009$~\cite{Workman:2022ynf}, which does not enter the fit, 
resulting in a pull value of $2.1$. 

\begin{figure}
\centering
\includegraphics[width=0.48\textwidth]{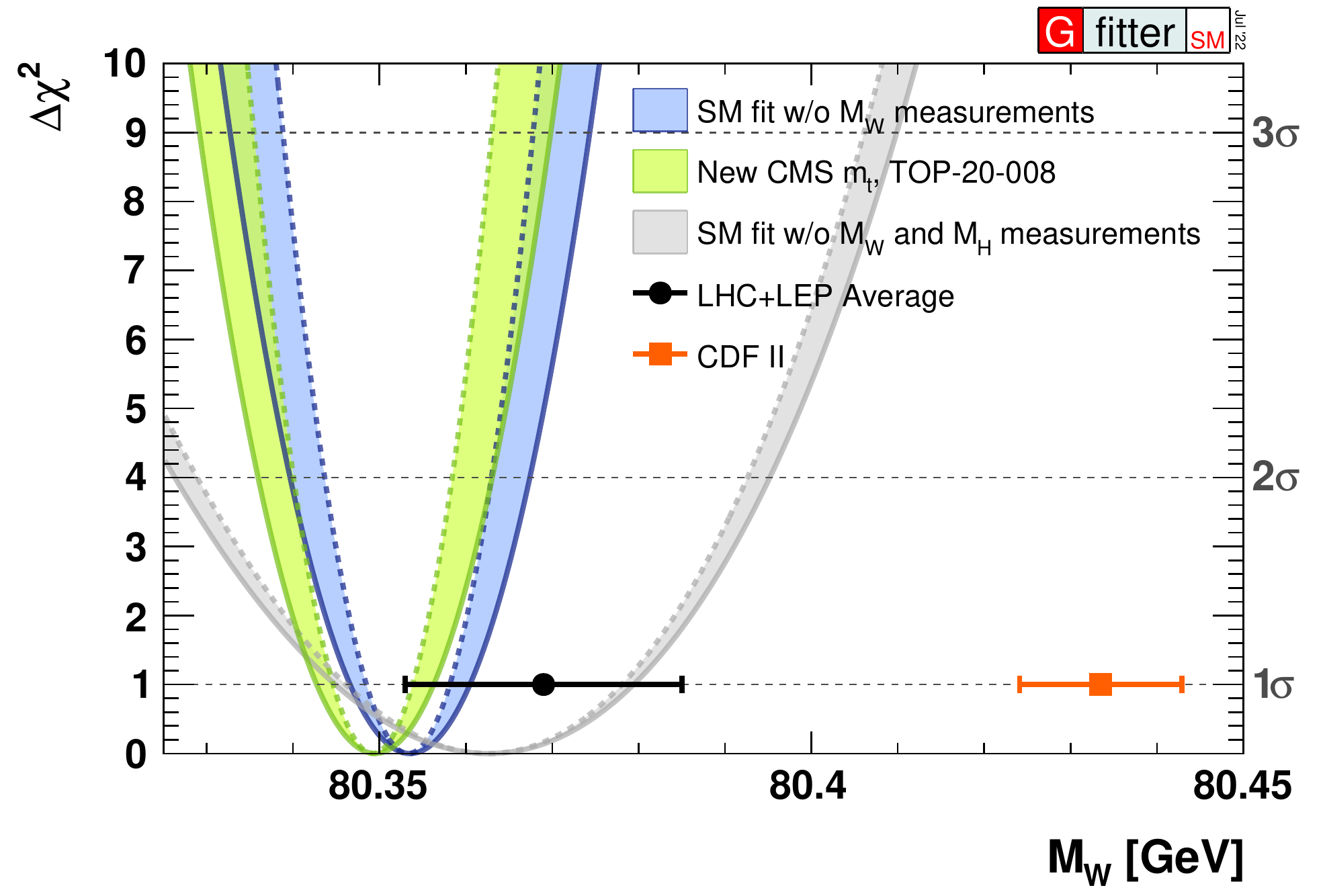}
\includegraphics[width=0.48\textwidth]{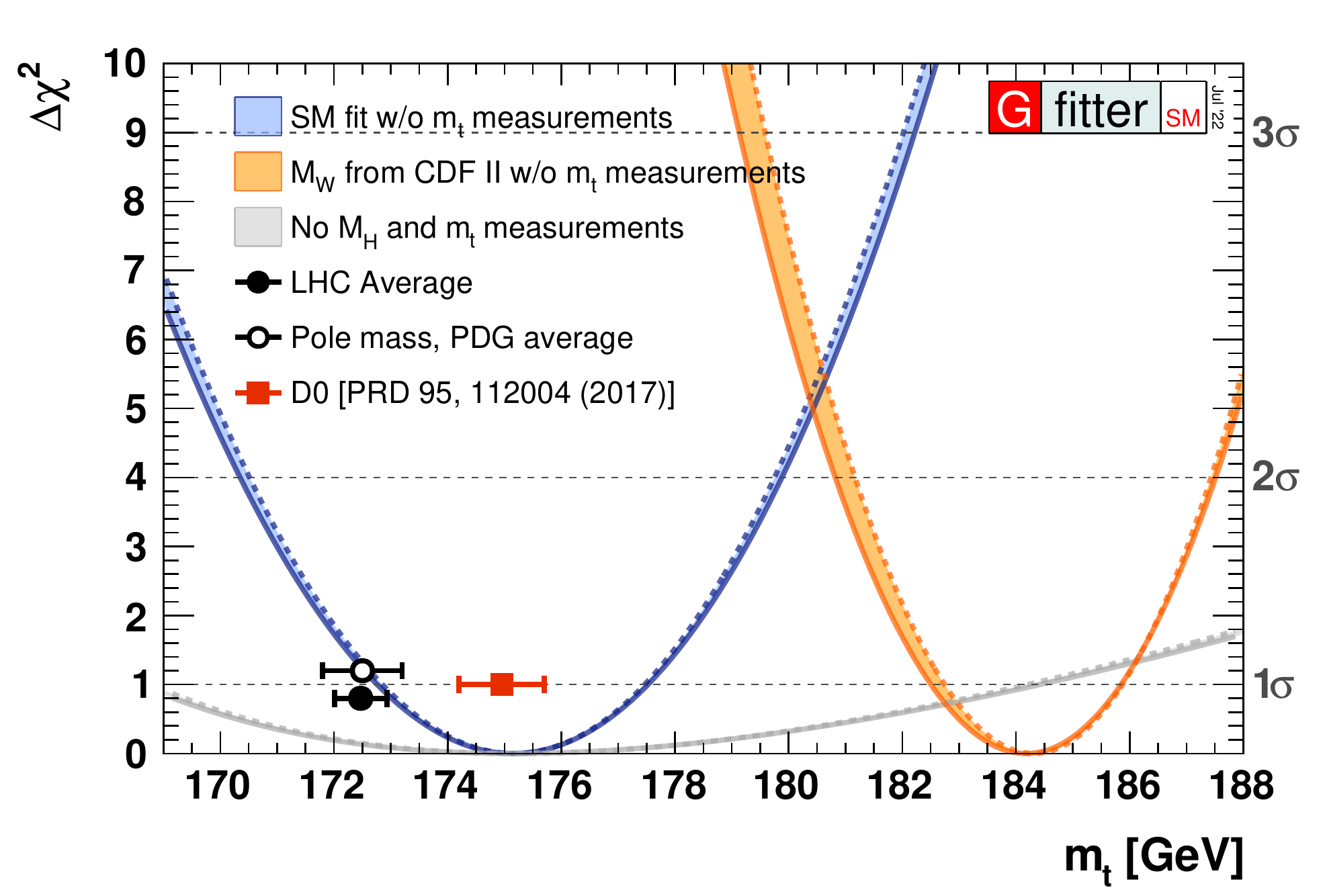}\hspace{0.1cm}
\caption[]{Results of the EW fit: Change in the negative log-likelihood 
function from its minimum ($\Delta\chi^2$), as a function of \MW (left) and \mt (right). 
The direct measurements of are excluded from the respective fits in these scans. \label{fig:fitmtmw}}
\end{figure}

The mass of the \PW boson shows a pull value of $-0.8$, which has decreased
with respect to our previous result, where it was $-1.5$. The reason for the better 
agreement with the prediction is the smaller value of \MW from our new combination. 
When not including \MW in the fit, we obtain a prediction of $\MW = 80354 \pm 7\mev$, 
with an uncertainty about half as large as the uncertainty in our \MW combination. 
A comparison between the measured values and this prediction 
is shown in Figure~\ref{fig:fitmtmw} (left). 
The disagreement between the \MW prediction and the CDF Run II value is $6.8\sigma$. 
It becomes larger when using the most recent CMS measurement 
of \mt~\cite{CMS-PAS-TOP-20-008} as input to the fit, shown as green band. 

We find good agreement between the measurements and predictions of \sinleff, with 
pull values of $-0.7$ and 0.5. The fit prefers a value of 
\mt which is close to the input value, where we find a best-fit value of $172.67\pm0.65\gev$. 
When not including \mt in the fit, we obtain a prediction of 
$\mt = 175.15^{\,+2.37}_{\,-2.39}\gev$, which is compatible with the direct measurements. 
If we assume perfect knowledge of \MW with a negligible uncertainty, the prediction 
of \mt has an uncertainty of 0.9\gev, emphasising the importance of precise 
\MW measurements.  
In Figure~\ref{fig:fitmtmw} (right), we compare the \mt prediction with direct 
measurements. Also shown is the prediction when using the CDF Run II \MW measurement 
instead of our combined value in the fit, resulting in $\mt = 184.2 \pm 1.7\gev$. 

The fit predicts $\MH = 100^{+25}_{-21}\gev$, which 
agrees with the measured value within $1\sigma$. When replacing our combined value 
of \MW with the CDF Run II \MW measurement, the fit predicts \MH to be 
$42.3^{+10.2}_{-8.7}\gev$. 
When removing all observables sensitive to \MH from the fit except for 
the CDF Run II \MW measurement, we find $\MH < 47\gev$ at 95\% confidence level. 

In general, when performing the full EW fit with the CDF Run II \MW measurement as 
input instead of our combined \MW, the fit results in a minimum 
$\chi^2/\mathrm{dof} = 62.6/15$, which corresponds to a $p$ value of 
$8.9 \cdot 10^{-8}$. The largest pulls are observed in \MW, \Afb{b} and \mt, 
with values of -5.5, 2.4 and 2.8, respectively.

\section{Summary}

We have presented the latest results of the global electroweak fit with Gfitter. 
The electroweak sector of the standard model of particle physics is tested to high precision, 
and it is found to be internally consistent with a $p$ value of 0.34. 
Naive combinations of \MW and \sinleff from the LEP, Tevatron and LHC experiments 
agree well with the SM predictions. We have shown that the new \MW measurement 
by the CDF Collaboration can not be incorporated in the SM fit without introducing 
large tensions.

\bibliography{References}{}

\end{document}

%% file: definitions.tex
\newcommand{\PW}    {\ensuremath{W}\xspace}
\newcommand{\PZ}    {\ensuremath{Z}\xspace}

\mathchardef\Upsilon="7107
\def\Y#1S{\ensuremath{\Upsilon{(#1S)}}\xspace}

\newcommand{\mt}{\ensuremath{m_{t}}\xspace}
\newcommand{\MW}{\ensuremath{M_{W}}\xspace}
\newcommand{\MH}{\ensuremath{M_{H}}\xspace}
\newcommand{\as}{\ensuremath{\alpha_{\scriptscriptstyle S}}\xspace}

\newcommand{\asZ}{\ensuremath{\as(M_Z^2)}\xspace}

\renewcommand\l{\ell}

\newcommand{\Afb}[1]{{\ensuremath{A_{\rm\scriptscriptstyle FB}^{#1}}}\xspace}

\newcommand{\Kbar    }{\kern 0.2em\overline{\kern -0.2em K}{}\xspace}

\newcommand{\Kz      }{\ensuremath{K^0}\xspace}
\newcommand{\Kzb     }{\ensuremath{\Kbar^0}\xspace}
\newcommand{\KzKzb   }{\ensuremath{\Kz \kern -0.16em \Kzb}\xspace}
\newcommand{\Kp      }{\ensuremath{K^+}\xspace}
\newcommand{\Km      }{\ensuremath{K^-}\xspace}

\newcommand{\KpKm    }{\ensuremath{\Kp \kern -0.16em \Km}\xspace}

\newcommand\Dbar    {\kern 0.18em\overline{\kern -0.18em D}{}\xspace}

\newcommand\Bbar    {\kern 0.18em\overline{\kern -0.18em B}{}\xspace}

\newcommand\Bz      {\ensuremath{B^0}\xspace}

\newcommand\Bzb     {\ensuremath{\Bbar^0}\xspace}
\newcommand\Bu      {\ensuremath{B^+}\xspace}
\newcommand\Bub     {\ensuremath{B^-}\xspace}

\newcommand\BpBm    {\ensuremath{\Bu {\kern -0.16em \Bub}}\xspace}
\newcommand\Bs      {\ensuremath{B^0_{s}}\xspace}
\newcommand\Bsb     {\ensuremath{\Bbar^0_{s}}\xspace}

\newcommand\BzBzb   {\ensuremath{\Bz {\kern -0.16em \Bzb}}\xspace}
\newcommand\BszBszb {\ensuremath{\Bs {\kern -0.16em \Bsb}}\xspace}

\newcommand{\tev}{\ensuremath{\,\mathrm{Te\kern -0.1em V}}\xspace}
\newcommand{\gev}{\ensuremath{\,\mathrm{Ge\kern -0.1em V}}\xspace}
\newcommand{\mev}{\ensuremath{\,\mathrm{Me\kern -0.1em V}}\xspace}
\newcommand{\kev}{\ensuremath{\,\mathrm{ke\kern -0.1em V}}\xspace}
\newcommand{\ev}{\ensuremath{\,\mathrm{e\kern -0.1em V}}\xspace}
\newcommand{\gevc}{\ensuremath{{\,\mathrm{Ge\kern -0.1em V\!/}c}}\xspace}
\newcommand{\mevc}{\ensuremath{{\,\mathrm{Me\kern -0.1em V\!/}c}}\xspace}
\newcommand{\gevcc}{\ensuremath{{\,\mathrm{Ge\kern -0.1em V\!/}c^2}}\xspace}
\newcommand{\mevcc}{\ensuremath{{\,\mathrm{Me\kern -0.1em V\!/}c^2}}\xspace}

\newcommand{\bei}{\begin{itemize}}
\newcommand{\eei}{\end{itemize}}
\newcommand{\beq}{\begin{equation}}
\newcommand{\eeq}{\end{equation}}
\newcommand{\beqn}{\begin{eqnarray}}
\newcommand{\eeqn}{\end{eqnarray}}
\newcommand{\beqns}{\begin{eqnarray*}}
\newcommand{\eeqns}{\end{eqnarray*}}
\newcommand{\bitm}{\begin{itemize}}
\newcommand{\eitm}{\end{itemize}}

\newcommand{\seffsf}[1]{\ensuremath{\sin\!^2\theta^{#1}_{{\rm eff}}}}

\newcommand{\sinfeff}{\seffsf{f}\xspace}
\newcommand{\sinleff}{\seffsf{\ell}\xspace}
